\documentclass[%
 aip,
 jcp,
 amsmath,amssymb,
 reprint,%
]{revtex4-1}

\usepackage{graphicx}
\usepackage{dcolumn}
\usepackage{bm}

\usepackage[utf8]{inputenc}
\usepackage[T1]{fontenc}
\usepackage{mathptmx}
\usepackage{etoolbox}
\usepackage{xcolor}
\usepackage{comment}
\usepackage[caption=false,subrefformat=parens,labelformat=parens]{subfig}
\usepackage{booktabs}
\usepackage{microtype}
\usepackage[all]{nowidow}
\makeatletter
\def\@email#1#2{%
 \endgroup
 \patchcmd{\titleblock@produce}
  {\frontmatter@RRAPformat}
  {\frontmatter@RRAPformat{\produce@RRAP{*#1\href{mailto:#2}{#2}}}\frontmatter@RRAPformat}
  {}{}
}%
\makeatother

\begin{document}

\preprint{AIP/123-QED}

\title[Breaking the mold]{Breaking the mold: overcoming the time constraints of molecular dynamics on general-purpose hardware}

\author{Danny Perez}
\affiliation{Los Alamos National Laboratory, Los Alamos, NM, USA}%

\author{Aidan Thompson}
\affiliation{Sandia National Laboratories, Albuquerque, NM, USA}%

\author{Stan Moore}
\affiliation{Sandia National Laboratories, Albuquerque, NM, USA}%

\author{Tomas Oppelstrup}
\affiliation{Lawrence Livermore National Laboratory, Livermore, CA, USA}%
\affiliation{Cerebras Systems, Sunnyvale, CA, USA}%

\author{Ilya Sharapov}
\affiliation{Cerebras Systems, Sunnyvale, CA, USA}%

\author{Kylee Santos}
\affiliation{Cerebras Systems, Sunnyvale, CA, USA}%

\author{Amirali Sharifian}
\affiliation{Cerebras Systems, Sunnyvale, CA, USA}%

\author{Delyan Z. Kalchev}
\affiliation{Cerebras Systems, Sunnyvale, CA, USA}%

\author{Robert Schreiber}
\affiliation{Cerebras Systems, Sunnyvale, CA, USA}%

\author{Scott Pakin}
\affiliation{Los Alamos National Laboratory, Los Alamos, NM, USA}%

\author{Edgar A. Leon}
\affiliation{Lawrence Livermore National Laboratory, Livermore, CA, USA}%

\author{James H. Laros III}
\affiliation{Sandia National Laboratories, Albuquerque, NM, USA}%

\author{Michael James}
\affiliation{Cerebras Systems, Sunnyvale, CA, USA}%

\author{Sivasankaran Rajamanickam}
\affiliation{Sandia National Laboratories, Albuquerque, NM, USA}%

\date{\today}

\begin{abstract}
The evolution of molecular dynamics (MD) simulations has been intimately linked to that of computing hardware.
For decades following the creation of MD, simulations have improved with computing power along the three principal
dimensions of accuracy, atom count (spatial scale), and duration (temporal scale). Since the mid-2000s, computer platforms
have however failed to provide strong scaling for MD as scale-out CPU and GPU platforms that provide substantial increases to spatial scale
do not lead to proportional increases in temporal scale. Important scientific problems therefore remained inaccessible to direct simulation, prompting the development of increasingly sophisticated algorithms that present significant complexity, accuracy, and efficiency challenges. While bespoke MD-only hardware solutions have provided a path to longer timescales for specific physical systems, their impact on the broader community has been mitigated by their limited adaptability to new methods and potentials.
In this work, we show that a novel computing architecture, the Cerebras Wafer Scale Engine, completely alters the scaling
path by delivering unprecedentedly high simulation rates up to 1.144M steps/second for 200,000 atoms whose interactions
are described by an Embedded Atom Method potential. This enables direct simulations of the evolution of materials using
general-purpose programmable hardware over millisecond timescales, dramatically increasing the space of direct MD simulations that can be carried out.
\end{abstract}

\maketitle
\section{\label{sec:introduction}Introduction}

Since its introduction in the late 1950s \cite{alder1959studies,Rahman1964}, molecular dynamics (MD) has established itself as one of the most powerful simulation workhorses in the computational physical sciences. By generating  unbiased trajectories that are fully atomistically resolved, MD in principle enables the calculation of any thermodynamic (thermodynamic averages, phase diagrams) or dynamic (transition rates for chemical reactions, out-of-equilibrium behavior such as shock loading) properties that depend on atomic degrees of freedom. This generality and formal simplicity makes MD extremely versatile, contributing to its permeation into many sub-fields of chemistry, physics, materials science, engineering, and biology, as attested by the numerous scientific papers that use or refer to MD simulations.

The history and evolution of MD naturally has been intimately tied with the evolution of the hardware on which the simulations are carried out. While the exponential increase in the capability of large-scale computing systems has brought wide-ranging scientific benefits through a dramatic expansion of the accessible MD simulation space (as defined by a combination of simulation accuracy, size, and time), the technological constraints inherent to microprocessor technology has been reflected into an extremely skewed growth. This ``co-evolution'' has deeply affected the practice of MD and consequently the scientific problems on which it has been brought to bear. This is an example of the ``hardware lottery'' phenomenon, a term that has been used in the field of artificial intelligence to describe how the characteristics of available hardware dictate which specific dimensions of a problem receive attention\cite{hooker2021hardware}.

\section{\label{sec:history}Co-evolution of MD and computing technologies}

As will be discussed below in more detail, growth in aggregate computing power can stem from multiple sources. Growth driven by the exponential increase in clock frequency experienced from the 1970s to the mid 2000s \cite{cpudb} led to a very ``isotropic'' expansion of the simulation space, as increases in the operation execution rate of a single processor with a small number of execution units can be productively channeled into arbitrary combinations of increased accuracy through the arithmetic intensity of the physical models used to approximate quantum-mechanical energies and forces  (the so-called interatomic potentials---IPs), of increased number of atoms simulated at fixed simulation time (weak-scaling), or of increased simulation times at fixed problem size (strong-scaling).

This ``golden age'' came to an end with the gradual shift toward increases in aggregate computing power being primarily delivered by scaling the number of processors integrated into a common communication fabric in the form of massively parallel distributed-memory machines. These architectures are natural fits to domain-decomposition approaches where different spatial sub-domains are mapped into different processors, enabling essentially limitless weak-scaling, fueling a continuous and dramatic increase in the size of the largest simulations that can be carried out. This era has witnessed milestone simulations, including the first trillion-atom simulation in 2008 \cite{germann2008trillion}. The rise of massively parallel architectures also led to more accurate IPs, at a cost in computational intensity. The ability to increase the amount of computation required per atom without increasing the communication burden enabled moderately large simulations to be distributed over entire petascale and now exascale supercomputers\cite{Trott2014, li2024scaling}.

However, the need for inter-domain communication at every MD time step, inherent to domain-decomposition approaches, essentially marked the end of the sustained increase in the time horizon accessible to direct simulation because the bandwidth and latency of the communication fabric then ultimately controlled the maximal sustainable time-stepping rate. Unfortunately, the performance of networking technologies has not evolved following its own Moore's Law growth curve, leading to a stagnation in the maximal simulation duration, and leading to MD being mostly understood as a \emph{sub-microsecond} timescale method.

In more recent years, the distributed-memory model has been augmented by the introduction of powerful GPUs that gradually replaced conventional CPUs as the main computational engines, delivering very favorable operation/watt performance. The massively data-parallel nature of GPUs proved a natural fit to the emergence of machine-learned IPs \cite{behler2016perspective,deringer2019machine}, whose very architecture often co-evolved with that of GPUs. This has driven the recent community-wide focus on dramatically increasing the accuracy of MD simulations through extremely arithmetically intensive energy and force calculations, with the aim of achieving ``near-quantum'' accuracy at an affordable cost. At lower accuracy, massively data-parallel GPU machines can still deliver weak-scaling, although with an even coarser granularity, as millions of atoms per GPU are now required to maintain a high hardware utilization and low communication overhead for the ``classic'' light-weight IP computations that dominated the first three decades of MD.

The historical technological evolution after the end of this golden age has led to an extremely skewed expansion of the MD simulation space, with profound scientific consequences. Heroic, very large MD simulations have enabled the direct exploration of complex mesoscopic physics that unfold on relatively short timescales, including classical instabilities such as Rayleigh-Taylor \cite{kadau2007importance}, the behavior of materials under shock loading \cite{swaminarayan2008369}, and the high-strain-rate plastic deformation of materials \cite{abraham2002simulating}, to name only a few. Scientifically, this has partially alleviated the need for complex phenomenological models that often necessitate the introduction of numerous approximations or heuristics that can be difficult to validate in practice or the time-consuming development of sophisticated multiscale/multiphysics models.

The widespread usage of high-fidelity IP models also has been supported by these technological developments, leading to the consideration of increasingly complex physical interactions and to the expansion of the space of materials that can be directly investigated from simple, pure materials to complex, multi-component molecular and condensed phase systems. These efforts have accelerated exponentially since the 2000s with the explosion of machine-learned IPs that leverage the increasing availability of reference quantum calculations to train extremely flexible models of atomic interactions \cite{behler2016perspective,deringer2019machine}. In this case also, the technological evolution of CPUs and GPUs makes them adept at tackling problems that exhibit high arithmetic intensity, driving rapid scientific progress.

The situation is completely different with respect to simulation times, the third and final axis of the simulation space. The stagnation of the accessible timescales caused by the saturation of clock frequencies has forced the community to develop an increasingly sophisticated arsenal of methods designed to circumvent the need for long trajectories. For example, the calculation of thermodynamic averages is plagued by metastability issues, where trajectories remain trapped in sub-regions of the full phase space for long periods of time, leading to extremely slow convergence of thermodynamic observables or worse, to pseudo-convergence to erroneous values. This has led to the development of a zoo of enhanced sampling methods \cite{rousset2010free,bernardi2015enhanced,yang2019enhanced} where convergence is accelerated by the introduction of various biases (e.g.,~as in umbrella sampling \cite{kastner2011umbrella}) and/or by the generation of large numbers of short trajectories (e.g.,~as in parallel tempering \cite{earl2005parallel}). These methods have proven extremely successful at improving the reliability of thermodynamics inferred from MD\@. However, by construction, they corrupt the natural dynamics of the system, making them inappropriate to study dynamical evolution over long timescales. Investigating these problems requires the introduction of even more sophisticated, dynamically accurate techniques \cite{elber2005long} (e.g.,~forward flux sampling \cite{allen2009forward}, accelerated molecular dynamics \cite{perez2009accelerated,zamora2020accelerated,perez2015parallel}, etc.). While these methods have also proven successful at providing unique insights into a broad range of chemical and physical systems (chemical reactions, microstructural evolution processes, protein folding, etc.), their use often requires deep domain knowledge of the target systems (e.g.,~through the knowledge of appropriate reaction coordinates) or loose computational efficiency for dynamically-complex systems, making investigation of the long-time properties slow and labor intensive, stifling progress on a range of fundamentally and technologically critical problems.

The divergence between the algorithmic needs of long-timescale MD and the technological constraints and market forces of the computing industry has motivated attempts at the development of custom hardware for MD simulations, such as the Anton \cite{anton_GB} and MDGRAPE \cite{susukita2003hardware} hardware families. These machines have proven successful at breaking the deadlock, enabling increases in simulation timescales by orders of magnitude, thereby enabling the direct simulation of complex biomolecular processes such as protein folding \cite{lindorff2011fast}. This dramatic improvement in performance however came at the cost of implementing a specific IP class in bespoke hardware, which is economically onerous and scientifically constraining, leading to spectacular successes for the systems they were designed to simulate but to an overall rather limited impact on the broader practice of MD outside of these specific communities.

\begin{figure}
\subfloat[][]{\includegraphics[width=\linewidth]{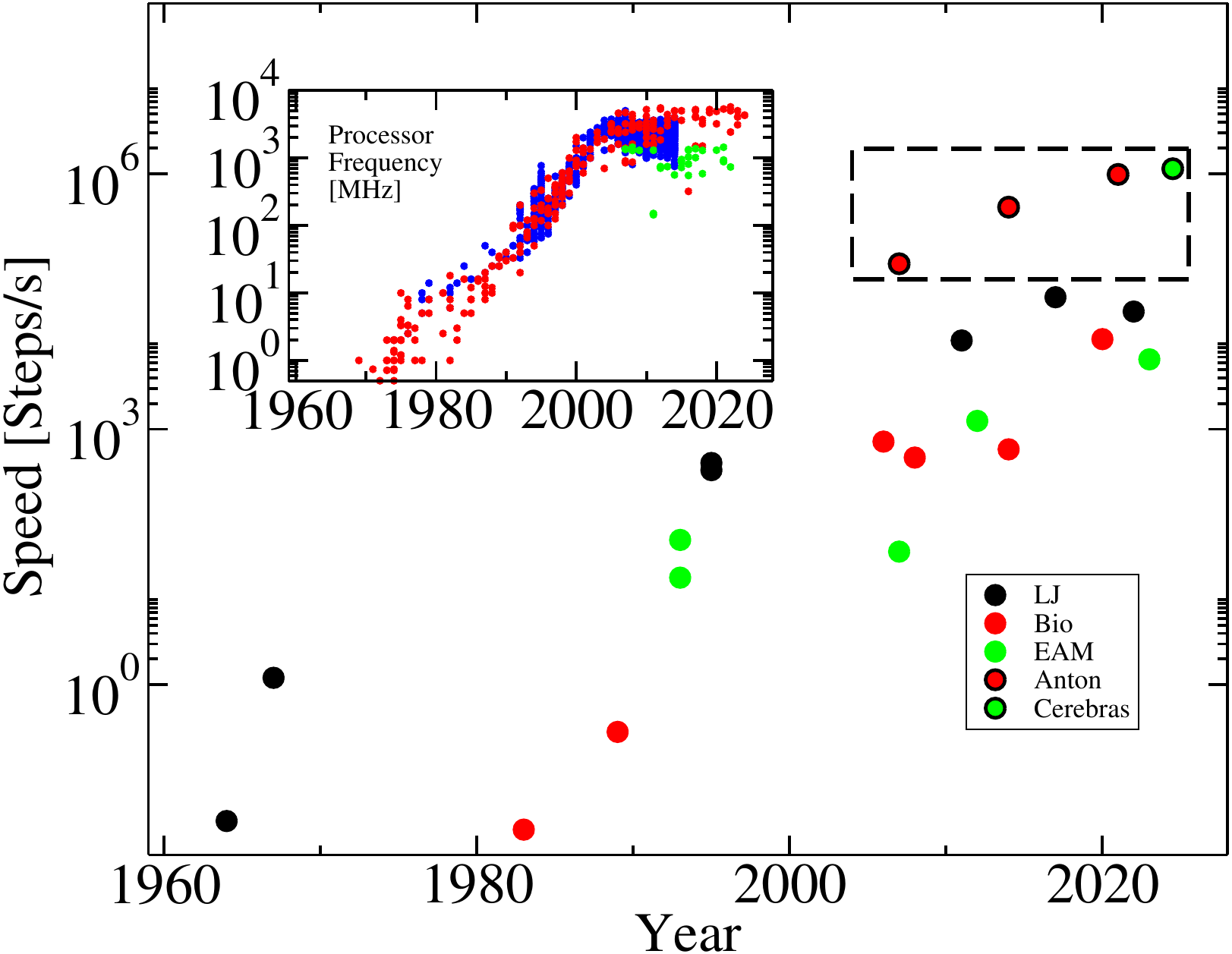}}\par
\subfloat[][]{\includegraphics[width=\linewidth]{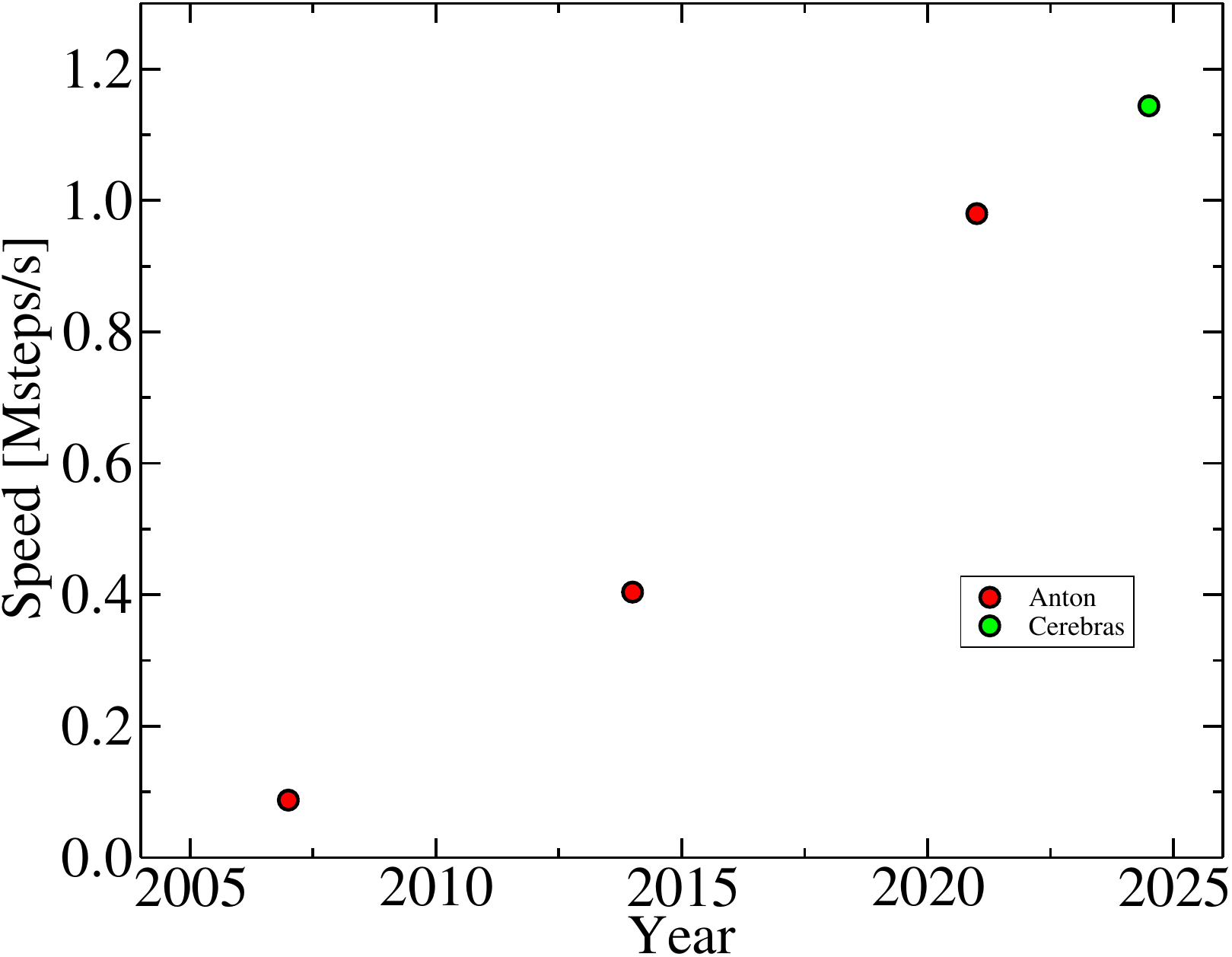}}
\caption{\label{fig:historystep} (a)~Progression of maximum simulation speed over six decades. Historically prominent benchmarks for several classes of interatomic interaction model are plotted versus time: Lennard-Jones (LJ, black), biomolecular (Bio, red) and Embedded Atom Method (EAM, green).  Benchmarks for all three model classes have stagnated since 2010. Exceptions to this are the Anton special-purpose ASIC supercomputers (red/black) and now the Cerebras wafer-scale engine (green/black). Inset: Historical clock frequency data for CPU processors (blue)\cite{cpudb}, (red)\cite{ cpuwikipedia} and GPU processors (green)\cite{gpgpuwikipedia}. (b)~An expanded view of Anton and Cerebras performance (Msteps/s, linear scale) during the time period of the dashed rectangle in~(a). More details on each of these benchmarks are provided in Table~\ref{tab:history}.}
\end{figure}

A more quantitative view of this narrative of exponential growth in simulation timescales followed by stagnation can be obtained by using maximum speed, measured in timesteps per second, as a proxy for timescale.  Table~\ref{tab:history} lists prominent milestones in maximum speed reported in the scientific literature.  We begin with Rahman's 1964 seminal paper that is celebrated in this Special Topic, where he reported that for 250 Lennard-Jones particles, ``each cycle takes 40 sec on the IBM-704 machine.''\cite{Rahman1964} It would be impractical to exhaustively sample the vast subsequent published literature on this topic, but we have attempted to record the fastest MD simulation speed reported in each era. From 1970 through 2005 the speed improvements closely tracked processor clock frequency, which increased by an order of magnitude about every 10 years.\cite{cpudb} Since then, clock speeds have stayed at or below 3~GHz due to physical limitations on power dissipation. This overall trend can be seen more clearly when the benchmark data is plotted versus time.  Fig.~\ref{fig:historystep} shows a set of historically prominent published benchmarks for several classes of interatomic interaction models, as well as clock frequency of processors.\cite{cpudb} Since 2010 there has been no significant improvement in maximum simulation speed.  Across a wide range of IPs, materials, codes, and hardware, the maximum speed saturates in the range $10^4$--$10^5$ steps/s. A notable exception to this are the Anton series of special-purpose machines custom-built for biomolecular simulations \cite{anton_GBlong}. There currently is an unmet need for machines that can run atomistic simulations of a wide range of materials and conditions at speeds greater than $10^6$ steps/s, opening the door to direct simulations over millisecond timescales that provide unprecedented insights into slow, thermally activated reactions and microstructural evolution.

\begin{table}[tp]
    \caption{Historical progression of maximum MD simulation speed from Rahman's seminal 1964 paper to today. The best result from the new work described in this paper is indicated in bold.}
    \label{tab:history}
    \centering
    \begin{tabular}{@{} r r r l r l l @{}}
      \toprule
      Year
      & \multicolumn{1}{r}{Step/s}
      & \multicolumn{1}{c}{$N$}
      & Model
      & \multicolumn{1}{c}{$P$}
      & Processor
      & Source \\
      \midrule
1964 & 0.025  & 250  & LJ  & 1    & IBM-704           & Ref.~[\!\citenum{Rahman1964}] \\
1967 & 1.2   & 864  & LJ  & 1    & CDC6600           & Ref.~[\!\citenum{Verlet1967}] \\
1983 & 0.0198 & 5208 & Bio  & 1 & Cyber170/760 & Ref.~[\!\citenum{Vangunsteren1983}] \\
1986 & 0.278 & 4069 & Bio  & 1 & Cyber205 & Ref.~[\!\citenum{Vangunsteren1986}] \\
1993 & 18    & 500  & EAM & 1    & YMP          & Ref.~[\!\citenum{Plimpton1993}] \\
1993 & 50    & 500  & EAM & 1024 & nCUBE2             & Ref.~[\!\citenum{Plimpton1993}] \\
1995 & 330   & 500  & LJ  & 1    & C90           & Ref.~[\!\citenum{Plimpton1995}] \\
1995 & 400   & 500  & LJ  & 1024 & Paragon      & Ref.~[\!\citenum{Plimpton1995}] \\
2006 & 714 & 24k & Bio  & 512 & Opteron & Ref.~[\!\citenum{Bowers2006}] \\
2007 & 36.4 & 386M & EAM  & 213k   & BG/L & Ref.~[\!\citenum{Glosli2007}] \\
2007 & 87,500 & 24k & Bio  & 512   & Anton-1 & Ref.~[\!\citenum{Shaw2008}] \\
2008 & 463 & 24k & Bio  & 128 & Xeon X5160 & Ref.~[\!\citenum{Hess2008}] \\
2011 & 11,000 & 4k & LJ  & 16   & Xeon E5-2670 & Ref.~[\!\citenum{Thompson2013}] \\
2012 & 1,250 & 2k & EAM  & 12   & Xeon X5690 & Ref.~[\!\citenum{lammpsbench}] \\
2014 & 579 & 32k & Bio  & 4096   & Xeon E5-2670 & Ref.~[\!\citenum{Moore2014}] \\
2014 & 404,000 & 24k & Bio  & 512   & Anton-2 & Ref.~[\!\citenum{Shaw2014}] \\
2017 & 35,400 & 512 & LJ  & 1 & Gtx1080 & Ref.~[\!\citenum{Bailey2017}] \\
2020 & 11,400 & 5k & Bio  & 1 & RTX2080 & Ref.~[\!\citenum{Pall2020}] \\
2021 & 980,000 & 24k & Bio  & 64   & Anton-3 & Ref.~[\!\citenum{anton_GB}] \\
2022 & 24,000 & 1k & LJ  & 36   & Skylake & Ref.~[\!\citenum{Thompson2022}] \\
2023 & 6,600 & 3.5M & EAM  & 1.9M & A64FX & Ref.~[\!\citenum{Li2023}] \\
2024 & 704,225 & 0.8M & EAM & 0.8M & WSE-2 & Ref.~[\!\citenum{santos2024breaking}] \\
\bfseries 2024 & \bfseries 1,144,000 & \bfseries 0.2M & \bfseries EAM & \bfseries 0.8M & \bfseries WSE-2 & \bfseries This work \\
      \bottomrule
    \end{tabular}
\end{table}

In this manuscript, we demonstrate that a new general-purpose hardware architecture---the Cerebras Wafer Scale Engine (WSE)---can overcome the traditional scaling constraints that have shaped the practice of MD for the last two decades, allowing access to uncharted regions of the MD simulation space while retaining the programmability of conventional architectures. These advances could have wide-ranging impact on many problem areas that require understanding of the long-time behavior of materials. Such problems are pervasive, as the complex structure of potential energy landscapes of most materials commonly results in strongly metastable dynamics where the key evolution processes (e.g.,~chemical reactions, molecular folding, defect nucleation, diffusion, reactions, or topological relaxation in glasses) occur on timescales that are much longer than typical MD integration timesteps of a few femtoseconds \cite{Wales_2004}.

The manuscript is organized as follows.  The basic features of Cerebras' WSE architecture are summarized in Sec.~\ref{sec:wse}\@. Sec.~\ref{sec:methods} describes the approach behind the implementation. The performance analysis of the implementation and the comparison with a conventional MD implementation running on the Frontier supercomputer at the Oak Ridge Leadership Computing Facility (OLCF) are presented in Sec.~\ref{sec:results}\@.  A short perspective on the potential impact of this new capability is offered in Sec.~\ref{sec:discussion}, and Sec.~\ref{sec:conclusions} summarizes the results presented and draws some conclusions.

\section{\label{sec:wse}Wafer-Scale Engine}
We implemented an MD algorithm on a Cerebras Wafer-Scale Engine (WSE)\@.  WSEs are monolithic processors constructed as the largest square possible within a (circular) 300~mm silicon wafer. These 46,225~mm\textsuperscript{2} chips are the world's largest and also the most powerful in terms of peak performance. We outline aspects of the processor relevant to this work and refer readers to previous publications~\cite{WSE_HPC_2,HotChips} for additional details.

A WSE comprises a Cartesian grid of \emph{tiles}, each with a dedicated general-purpose processor core, dataflow fabric router, and primary memory. The WSE-2 processor, a second-generation WSE, contains 850,000 cores arranged in a ${\sim}920{\times}920$ array and draws 23~kW of power. Tiles are mesh-connected to their four nearest neighbor tiles on the wafer. Cores act independently of each other and can execute different code.

The WSE-2 provides a total of 40~GB of single-cycle latency, on-chip SRAM distributed across tiles (48~kB per tile). The cumulative memory bandwidth is 20~PB/s.  At the tile level, memory speed matches the speed of the computational core, avoiding the memory-access bottleneck.

The fabric routers can exchange up to ten 32-bit messages on every cycle: one in each direction to the local core and to all four neighboring routers. The latency between neighboring routers is one cycle.  The hardware supports 24 virtual channels on all physical links. In aggregate, the fabric delivers 20~PB/s of interconnection bandwidth matching the aggregate memory bandwidth.

The design of the WSE removes the need and the associated overhead of an operating system and a message-passing library like MPI~\cite{mpi41}. The tile's processor has native instructions to send and receive messages, with routing, queuing, and flow control done in hardware. The architectural elements open opportunities for novel algorithms and methods to speed up scientific calculations. Previous studies~\cite{WSE_HPC_1,WSE_HPC_2,WSE_HPC_3} demonstrate that the WSE is capable of delivering orders of magnitude higher performance than conventional systems for scientific workloads.

\section{\label{sec:methods}Method}

\begin{figure}[tp]
\centering
\subfloat[][\label{fig:t-shape2}]{\includegraphics[width=0.4\linewidth]{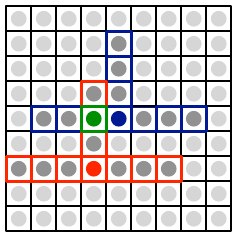}}\hfil
\subfloat[][\label{fig:diagonal}]{\includegraphics[width=0.4\linewidth]{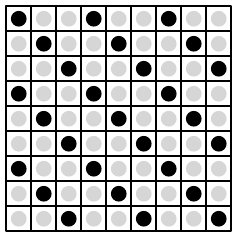}}
\caption{Optimizations of communication patterns that enable fine-grain parallelism. (a)~T-shaped communication pattern uses multicast along three one-dimensional segments. (b)~Midpoint core handling of interactions between two atoms. Black cores contain atoms, while gray cores are dedicated to processing atom interactions.}
\label{fig:mapping}
\end{figure}

In this work, we further develop a parallel algorithm for simulating atoms interacting through the Embedded Atom Method (EAM) \cite{daw1993embedded} IP\@. Our algorithm is designed specifically for processors connected by a 2D Cartesian mesh network. The following summarizes the design and implementation of this algorithm on the WSE platform.

Under EAM IPs, the system energy is a function of pairwise interactions and of an approximation of the embedding energy of an atom into the ambient electron density.  The system energy $U$ for $N$ atoms at positions $r_i$, $i=1,2,\ldots,N$, can be written as
\begin{equation}
  U = \sum_{i=1}^N \left\{ \sum_{j=i+1}^N \phi(|r_i - r_j|) + F\left(\sum_{j=1}^N \rho(|r_i - r_j|)\right)\right\} \;.
\end{equation}

The functions $\phi$ and $\rho$ are short-ranged and identically zero beyond some cut-off radius, typically a few atomic spacings. In efficient implementations, this fact can be leveraged to limit the ranges of the inner sums over $j$ above by using neighbor lists and spatial decomposition. In practical implementations, two data exchanges are performed: one to determine the density ($j$-sum over $\rho$) for each atom, and one to compute the $j$-sum over $\phi$ and assemble the forces from the $\rho$ values.

Here we consider a common EAM parameterizations for tantalum (Ta) \cite{li2003embedded} and tungsten (W) \cite{zhou2001atomic}. For the ground-state crystal structures, per-atom interaction counts are 14 for BCC Ta and 59 for BCC W\@.  (BCC = body-centered cubic.)

Our algorithm is an extension of Santos et~al.'s work \cite{santos2024breaking} that further increases strong scaling by assigning \emph{multiple processor cores per atom}. Assuming that processor cores are organized logically in a 2D Cartesian grid, atoms are mapped to the cores using spatial sorting for which the atomic coordinates are projected to a 2D grid by removing the $z$~coordinates. Rectangles of $nx \times ny$ cores are allocated to each 2D cell, and the $z$~column of atoms projected onto the cell are distributed over the corresponding core rectangle.

This spatial mapping ensures that there is a reasonably small bound on how far atom coordinates need to be communicated across the mesh network to compute forces in a way that captures all interatomic interactions.  We use a neutral territory method\cite{Midpoint,ForceDecomp1,ForceDecomp2} to determine the communication pattern and to assign work to processor cores. For each core, we broadcast its atom(s) along a T-shape (see Fig.~\ref{fig:mapping}\subref{fig:t-shape2}). The length of the arms of the T are determined so as to guarantee that all interacting pairs of atoms are found, i.e.,~that the coordinates for the atoms in each interacting pair are found on at least one core. It is close to optimal in minimizing the number of communication partners. Furthermore, it is particularly efficient for a 2D mesh network because data is transmitted only along directly connected cores and the aggregate network link bandwidth use is minimized.

The WSE's low-latency and high-bandwidth cross-core communication enables very fine-grained parallelism. Using the algorithm above and the fact that there are nearly a million cores on the WSE, we can allocate one atom per core or spread the problem even thinner by splitting each atom's interactions among multiple cores. In this case, the maximum size of the simulated system is proportionally scaled down as the number of cores per atom increases.

In our simulation experiments, we assign atoms to cores on the diagonals of the core grid. The spacing between the diagonals, $h$, determines the degree of parallelism (Fig.~\ref{fig:mapping}\subref{fig:diagonal}). The special case with $h=0$ (zero spacing) corresponds to the one-to-one mapping between atoms and cores. This diagonal assignment supports a particularly efficient use of the WSE's network by using a native multicast operation to transmit atom data to the $h$ adjacent unoccupied cores.

At the beginning of the simulation and at regular intervals, the possible interactions assigned to a core are screened to determine which are in fact within the cutoff distance (plus some skin thickness), and a neighbor list is built to avoid unnecessary computations. After each data exchange, the neighbor list is used to pack the relevant received data into contiguous memory to allow efficient vector operations on the data in the force calculation steps.

In order to maintain an efficient mapping of atoms as they move during a long simulation, a greedy remapping step is performed periodically. The procedure uses two neighborhood exchanges. First, cores exchange atom state and calculate the change in assignment cost for all swaps they could participate in. Then, cores exchange the identifier of their best swap partner. When a core detects a mutual agreement of swap preference, it overwrites its local atom state. In our experiments, this remapping step is infrequent enough that it exhibits only a small impact on the overall simulation performance.

We have used the implementation discussed above to perform comparative benchmark simulations on the ground state crystal phase for Ta and W at ambient conditions using the potentials referenced above on the WSE\@. For these benchmarks we used geometries akin to what one might use to study grain-boundary evolution, which can involve transformation processes on microsecond to millisecond timescales and require prohibitively many time steps on conventional computers. These geometries are thin slabs with 6 unit cells in the $z$~direction and dozens to hundreds of unit cells in the $x$ and $y$~directions, with atom counts ranging from $\sim$133,000 to $\sim$800,000, depending on the number of atoms per core used in the calculations. The results are summarized in Table~\ref{tab:perf-wse} and discussed further in Sec.~\ref{sec:results}.

We also ran benchmarks on the OLCF Frontier exascale supercomputer. Each Frontier node consists of 8 AMD Instinct MI250X graphics compute dies (GCDs) attached to an AMD Optimized 3rd Generation EPYC 64C 2GHz CPU. A total of 9,408 nodes are connected together by HPE's Slingshot-11 network. Further details on Frontier can be found at \url{https://www.top500.org/system/180047/}.

We used the LAMMPS~\cite{LAMMPS} KOKKOS package, which implements performance portability abstractions from the Kokkos library~\cite{kokkos,kokkos2}, to run on the AMD GPUs using the HIP backend.
GPU-aware MPI was used for multi-GPU runs on Frontier.

We used FP32 (single precision) on WSE because the WSE lacks FP64 (double precision) hardware. Simulations on Frontier used FP64 because the LAMMPS KOKKOS package does not yet support either FP32 or mixed FP64 and FP32. We note that Frontier's MI250X GPU has equal theoretical peak compute performance for FP32 and FP64.

For Fig.~\ref{fig:historytimeatoms}, short benchmarks for EAM were run on 128 Frontier nodes (1024 GCDs) and then atom counts were ideally weak-scaled up to the full machine node count (9408 nodes) and simulation time was ideally scaled up to 24 hours of wall-clock time. Single GCD benchmarks on Frontier were ideally scaled to 24 hours of wall-clock time but were not scaled in size. Simulations results from WSE, Fugaku, and Anton were also scaled to 24 hours of wall-clock time. Simulations on Fugaku used an EAM potential for copper \cite{copper2}, while our simulations on WSE and Frontier used EAM tantalum \cite{tantalum}.

A quasi-2D geometry with fully open (non-periodic boundary conditions) was used on Frontier and WSE to simulate a perfect crystal with no defects at room temperature (300 K) and at the equilibrium density. The $z$ dimension was held fixed at 6 replications of the unit cell, while the $x$ and $y$ dimensions were scaled to give the desired number of atoms. The quasi-2D geometry was chosen to mimic a grain boundary evolution problem \cite{santos2024breaking}.


\begin{table}
    \caption{Measured performance on the WSE platform for two different variants of EAM\@. Conditions leading to the maximal simulation speed is highlighted in bold.}

    \label{tab:perf-wse}
    \centering

    \begin{tabular}{l r c r r}
      \toprule
      Element & Cores/atom & N$_\mathrm{atoms}$ &   Steps/s \\
      \midrule
      tantalum (Ta)
      & 1 & 800,000 & 700,000 \\
      & 2 & 400,000 & 912,000 \\
      & 3 & 266,666 & \phantom{00}1,000,000 \\
      & \bfseries 4 & \bfseries 200,000 & \bfseries \phantom{00}1,144,000 \\
      & 5 & 160,000 &\phantom{00}1,020,000 \\
      & 6 & 133,333 & \phantom{00}1,060,000 \\

      \addlinespace

      tungsten (W)
      & 1 & 800,000 & 314,000 \\
      & 2 & 400,000 & 364,000 \\
      & 3 & 266,666 & 378,000 \\
      & 4 & 200,000 & 416,000 \\
      & 5 & 160,000 & 426,000 \\
      & 6 & 133,333 & 503,000 \\
      \bottomrule
    \end{tabular}
\end{table}

\section{\label{sec:results}Results}

\begin{figure}
\includegraphics[width=\linewidth]{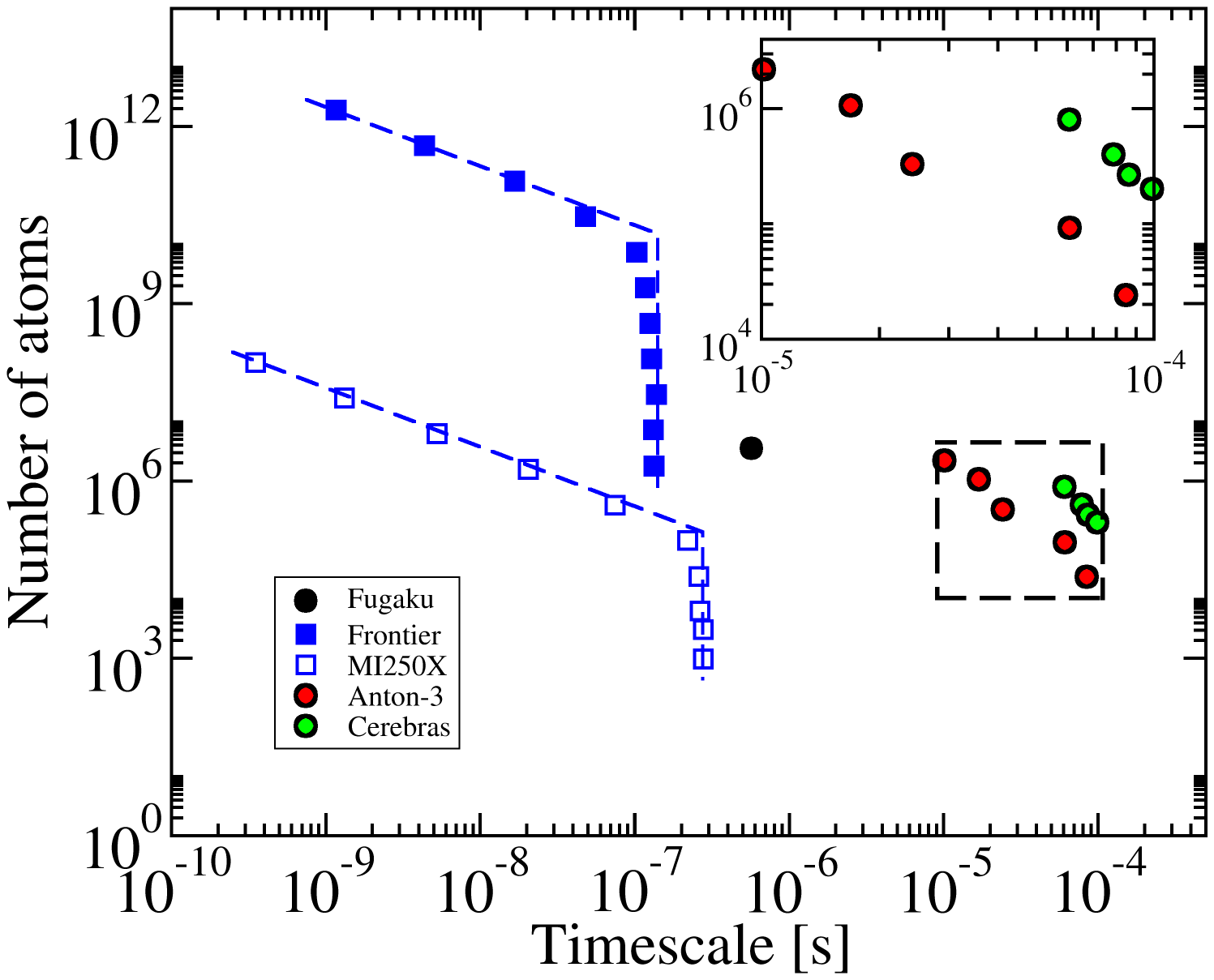}
\caption{\label{fig:historytimeatoms}  Accessible simulation space, assuming 24 hours of run time with 1~fs timestep size for EAM models running on Fugaku (black), Frontier (blue), and Cerebras (green/black), and a biomolecular model on the 64 node Anton-3 supercomputer (red/black). For Frontier, we assume that the small-scale benchmarks (obtained on 128 nodes) can be weak-scaled to the full machine (9,408 nodes). We also show the performance of small atom counts on a single Frontier MI250X GCD\@. Dashed lines indicate upper bounds on performance for Frontier and MI250X in the limits of ideal scaling (diagonal) and maximum speed (vertical).  Inset: Enlarged view of the Anton-3 and Cerebras points within the dashed rectangle.
}
\end{figure}

Fig.~\ref{fig:historytimeatoms} compares our EAM implementation's performance with that of state-of-the-art implementations on a few other computational platforms. Benchmarks carried out on OLCF's Frontier, the first exascale supercomputer, show a saturation in performance below 0.3\textmu{}s per (wall-clock) day, which is achieved for small systems on a single GCD, implying that inter-device communication is not required. As discussed above, Frontier is however ideally suited to weak-scaling.  It can simulate at very high parallel efficiency systems containing trillions of atoms, thereby delivering an unprecedented maximum simulation throughput of around 25 trillion atom-timesteps/second.  Simulation rates show a slight improvement on the Fugaku supercomputer, however remaining below 1\textmu{}s per day for a 3.5M atom system. Significant increases of simulation rates until now required special-purpose hardware such as the Anton family of platforms, which can reach very high simulation rates only slightly below 1M timesteps/second for relatively small systems (24k atoms), a performance level that has been unrivaled by general-purpose platforms for almost 20 years.

On the WSE, the results show that for a short cutoff potential (using EAM for Ta), simulation rates can reach \emph{1.144M timesteps per second} when using 4 cores/atom, which corresponds to about 0.1ms of simulated time per day. This is, as far as we know, the fastest reported MD simulation rate on any platform.  Aggressive parallelization with multiple cores per atom enables an additional speedup of about 63\% compared to the (already very aggressive) one core per atom baseline \cite{santos2024breaking}. This approach also enables a similar relative speed increase for the longer-ranged W~potential, providing a 60\% speedup when using 6 cores per atom. While the efficiency falls short of the ideal behavior, such extremely high simulation speeds can be  invaluable when studying problems where achieving long timescales is scientifically essential. It is also important to note that the arithmetic intensity of EAM is very low, which makes strong-scaling particularly difficult. Indeed, at 4 cores per atom, only 743 cycles are required to complete a timestep on the full system. Parallelizing more compute-intensive potentials using this approach could significantly improve efficiency at the cost of an overall lower simulation rate. Finally, although not currently implemented, a many-atoms-per-core version of the code is expected to allow for excellent weak-scaling, as the need for communication will be further reduced by improvements in data locality.

We note that the Anton architecture implements the CHARMM force field which relies on simpler pairwise potentials, in contrast to the many-body nature of EAM which requires two communication passes at every timestep. However, CHARMM also includes a long-range Coulombic contribution that is resolved every few timesteps, requiring additional computations and a more complex communication pattern. These qualitative differences make the detailed comparison of performance on Anton and WSE difficult.

\section{\label{sec:discussion}Discussion}


The unique combination of hardware platform and software implementation presented in this article addresses a critical regime that is inaccessible to conventional hardware/software combinations: simulating systems of moderate size ($10^5$--$10^6$ atoms) for timescales of \emph{0.1ms per simulation-day}.  These unprecedented timescales enable the direct investigation of complex slow-dynamics, including complex sluggish diffusive dynamics that does not exhibit any significant separation of timescales (which is prototypical of a dynamics that is extremely difficult to rigorously accelerate even with advanced methods), or to observe slow thermally-activated micro-structural evolution of materials directly in application-relevant conditions. As shown in Fig.~\ref{fig:historytimeatoms}, this critical regime is inaccessible to conventional massively-distributed platforms, which are limited to simulation rates on the order of 0.3\textmu{}s per simulation-day---three orders of magnitude less than what our work achieved.  Even more importantly, this regime is expected to \emph{remain} inaccessible even when extrapolating the performance of massively-distributed platforms based on vendor roadmaps.

Generalizing our approach to other potentials is ongoing work. The WSE's fast communication fabric permits efficient calculations even of the long-range interactions needed by biomolecular and polymer simulations. While the focus of this paper is on a dramatic extension in the time dimension of the accessible MD simulation space, incorporating ML potentials in our approach will expand the simulation space in both directions of time and accuracy. Furthermore, the Cerebras architecture is designed to combine efficiently into clusters of WSEs. Our preliminary modeling indicates that using multiple WSEs in a cluster largely maintains the time-stepping rate observed in this work, while substantially extending the accessible problems sizes. Consequently, general-purpose wafer-scale architectures like the WSE promise significant increase in all dimensions of the simulation space, opening opportunities for exploration and disruptive new discoveries in multiple fields that employ MD.

\section{\label{sec:conclusions}Conclusions}

The reach and scientific impact of molecular dynamics have been tied intimately to the evolution of the hardware platforms on which MD simulations are executed. While hardware evolution until around 2005 has led to largely transferable improvements across all three axes of the simulation space---length, time, and accuracy---the saturation of clock speeds since then has led to a very lopsided expansion of the simulation space. Indeed, massively-distributed architectures are particularly well-suited to domain-decomposition approaches that excel at weak-scaling, leading to a dramatic extension of accessible length scales. The introduction of GPUs also tends to translate into good performance for codes exhibiting high arithmetic intensity and hence benefited the development of more sophisticated interatomic potentials.  This includes the introduction of machine-learning-based models.

In contrast to increased length scales, simulation timescales have stagnated, as strong-scaling on massively-parallel platforms becomes limited by the latency of the communication fabric, the latency of launching kernels on the GPU, and other factors.  This saturation in simulation timescales has had significant scientific consequences, requiring the development of sophisticated enhanced sampling or long-time dynamical simulation methodologies that introduce substantial complexity and/or approximations that significantly hamper the investigation of long-time properties of materials.

Specialized hardware designs such as Anton and MDGRAPE have shown that timescale limitations can be substantially alleviated, although these approaches are rigid in model form and require considerable time and monetary investments, thereby restricting their broad scientific impact. In this work we demonstrate that dramatic extensions in the simulation space of MD applications are possible by leveraging novel fully-programmable architectures coupled with novel implementations, especially with respect to directly accessing long simulation times. Specifically, we leveraged Cerebras' WSE-2 processor, which delivers high performance by integrating a large computational resource in a single chip, balancing the speed of compute, memory, and communication. This hardware, combined with our EAM implementation that is optimized for it, dramatically curbs a long-standing historical trend that has hampered scientific discovery over the last 20 years, enabling direct fully-atomistic simulations over millisecond timescales. This approach is in the process of being generalized to other models of interatomic interactions and hence is expected to have a broad impact on the future development and practice of molecular dynamics.

\section*{\label{sec:acknowledgements}Acknowledgements}
Sandia National Laboratories is a multimission laboratory managed and operated by National Technology and Engineering Solutions of Sandia, LLC., a wholly owned subsidiary of Honeywell International, Inc., for the U.S. Department of Energy’s National Nuclear Security Administration under contract DE-NA-0003525. The SNL document release number is SAND2024-15616O. Los Alamos National Laboratory is operated by Triad National Security, LLC, for the National Nuclear Security Administration of U.S. Department of Energy (Contract No. 89233218CNA000001). The LANL document release number is LA-UR-24-32162. This work was performed under the auspices of the U.S. Department of Energy by Lawrence Livermore National Laboratory under Contract DE-AC52-07NA27344. The LLNL document release number is LLNL-JRNL-2000971.
The authors thank Ryan Humble for his contributions to the initial implementation specification.
This research used resources of the Oak Ridge Leadership Computing Facility at the Oak Ridge National Laboratory, which is supported by the Office of Science of the U.S. Department of Energy under Contract No. DE-AC05-00OR22725.

\section*{References}

\bibliography{references}

\end{document}